\documentclass[
reprint,
amsmath,amssymb,
superscriptaddress,
pra,
]{revtex4-2}

\usepackage{graphicx}
\graphicspath{{.}} % Path to figures folder
\usepackage{dcolumn}
\usepackage{bm}
\usepackage{physics}
\usepackage{braket}
\usepackage{placeins}
\usepackage[colorlinks=true]{hyperref}
\usepackage{upgreek}
\usepackage{multirow}
\usepackage{dsfont}
\usepackage{color,soul}
\begin{document}

\let\oldaddcontentsline\addcontentsline% Store \addcontentsline
\renewcommand{\addcontentsline}[3]{}% Make \addcontentsline a no-op
\newcommand{\citesec}[2]{\cite[\S{}#2]{#1}}% To reference supp. material

\title{
    Microwave-driven same-species sympathetic cooling for trapped ions\\
}

\begin{abstract}
    Sympathetic cooling of data qubits by coolant ions is an essential technique for trapped-ion quantum computing. 
    Conventionally a second ion species is used, requiring additional lasers and complicating optical setups. 
    We propose a scheme for sympathetic cooling using the same species and test it for $^{43}$Ca$^+$ ions. 
    Pulsed sideband cooling and ion addressing are implemented via integrated microwave control, further simplifying optical requirements. 
    We cool a two-ion gate mode close to its ground state ($\bar{n}\approx 0.16$) and benchmark an induced error on the data qubit of $1.7(4)\times 10^{-4}$ per cooling cycle.
 \end{abstract}

\author{M.\,C.\,Smith}
\thanks{These authors contributed equally to this work.}
\affiliation{Clarendon Laboratory, Department of Physics, University of Oxford, Parks Road, Oxford OX1 3PU, U.K.}

\author{E.\,Vandrey}
\thanks{These authors contributed equally to this work.}
\affiliation{Clarendon Laboratory, Department of Physics, University of Oxford, Parks Road, Oxford OX1 3PU, U.K.}

\author{A.\,D.\,Leu}
\affiliation{Clarendon Laboratory, Department of Physics, University of Oxford, Parks Road, Oxford OX1 3PU, U.K.}

\author{N.\,Drotleff}
\altaffiliation[Current address: ]{Institute for Quantum Electronics, ETH Z\"urich, Otto-Stern-Weg 1, 8093 Z\"urich, Switzerland}
\affiliation{Clarendon Laboratory, Department of Physics, University of Oxford, Parks Road, Oxford OX1 3PU, U.K.}

\author{A.\,Agrawal}
\affiliation{Clarendon Laboratory, Department of Physics, University of Oxford, Parks Road, Oxford OX1 3PU, U.K.}

\author{K.\,Miyanishi}
\altaffiliation[Current address: ]{Qubitcore Inc., OIST Innovation Core2 OIC2 207 1919-1 Tancha, Onna-son, Kunigami-gun Okinawa 904-0495, Japan}
\affiliation{Clarendon Laboratory, Department of Physics, University of Oxford, Parks Road, Oxford OX1 3PU, U.K.}
\affiliation{Center for Quantum Information and Quantum Biology (QIQB),\\
The University of Osaka, 1-2 Machikaneyama, Toyonaka 560-0043, Japan}

\author{D.\,M.\,Lucas}
\affiliation{Clarendon Laboratory, Department of Physics, University of Oxford, Parks Road, Oxford OX1 3PU, U.K.}

\author{M.\,F.\,Gely}
\affiliation{Clarendon Laboratory, Department of Physics, University of Oxford, Parks Road, Oxford OX1 3PU, U.K.}

\date{\today}

\maketitle

%!TEX root = output/arxiv.tex

%%%%%%%%%%% CONTEXT
%
Trapped ion qubits are one of the most promising approaches to quantum information processing, and have been used to demonstrate gate fidelities~\cite{smith2025,srinivas2021,clark2021,hughes2025} and state-preparation and measurement~\cite{sotirova2024,an2022,ransford2021,harty2014} well beyond error correction thresholds~\cite{campbell2017}.
Schemes to drive entangling gates generally use a shared motional mode of ions trapped in the same potential well as a ``quantum bus'' for distributing quantum information~\cite{cirac1995,molmer1999,sorenson2000}.
Some entangling gate schemes are designed to reduce the sensitivity to imperfect ground-state cooling, however their gate errors still increase with the temperature of the motional mode~\cite{molmer1999,sorenson2000,sutherland2024,sutherland2022,hughes2025}.
Heating of the motional modes can result from electric field noise~\cite{brownnutt2015}, or transport of ions in a quantum charge-coupled device (QCCD) architecture~\cite{kielpinski2002,moses2023}.
Long sequences of gates with low error rates therefore require a scheme which cools the motional mode whilst preserving the quantum information stored in the ions' electronic states.
This can be achieved via sympathetic cooling~\cite{kielpinski2000,rohde2001,barrett2003,rosenband2007}, where ancillary ``coolant'' ions are used to cool the shared motional modes of a trapped ion crystal without affecting information stored in the electronic states of the ``data'' ions.
%

%%%%%%%%%%% STATE OF THE ART (SYMPATHETIC COOLING)
%
Sympathetic cooling of trapped ions is therefore essential for quantum computing~\cite{home2009,blinov2002,barrett2003}, but also finds applications in precision spectroscopy~\cite{rosenband2007,imajo1996,roth2005,gruber2001,schiller2007,drewsen2007}, forming and cooling of molecular ions~\cite{roth2006,roth2006_spectroscopy}, and simulating chemical reactions~\cite{willitsch2008,gingell2010,staanum2008}.
A common approach is to use two different ion species~\cite{morigi2001,barrett2003,rosenband2007,imajo1996,roth2005,gruber2001,schiller2007,drewsen2007} which have a large frequency difference between their electronic transitions, such that lasers used to cool one species cause negligible photon scattering errors on the other.
However, one then requires the laser infrastructure to address additional wavelengths in the coolant ion.
Furthermore, for certain modes, the mass mismatch between ion species leads to decreased mode participation of the coolant ion, meaning indirect cooling schemes must be employed for efficient cooling~\cite{hou2024}.
These issues can be mitigated by using two isotopes of the same element~\cite{home2009,blinov2002}.
However, the smaller isotope shift between between the isotopes' atomic transitions can lead to larger off-resonant scattering errors, which must be reduced by spatially-addressing only the coolant ion, or limiting the number of scattered photons by using only sideband cooling rather than Doppler cooling.
If using the same isotope for both data and coolant ions, quantum information can be temporarily or permanently ``hidden'' in metastable states; however this exposes the qubit to imperfect mapping pulses and to decay to the ground state~\cite{allcock2021}.
Finally, if the coolant and data ions are the same isotope and the data qubit is encoded within the electronic ground-state manifold, quantum information can still be protected through the use of individual addressing techniques to target only the coolant, provided only sideband cooling (rather than Doppler cooling) is used to limit photon scattering errors.
Past demonstrations have used tightly-focused lasers to drive sidebands of only the coolant ions' quadrupole transition~\cite{rohde2001} or a local magnetic field gradient to provide frequency seperation between a coolant and data transition~\cite{sriarunothai2018}; however the ability to preserve or processs quantum information remains to be shown.

\begin{figure}[t]
    \centering
    \includegraphics[width=0.45\textwidth]{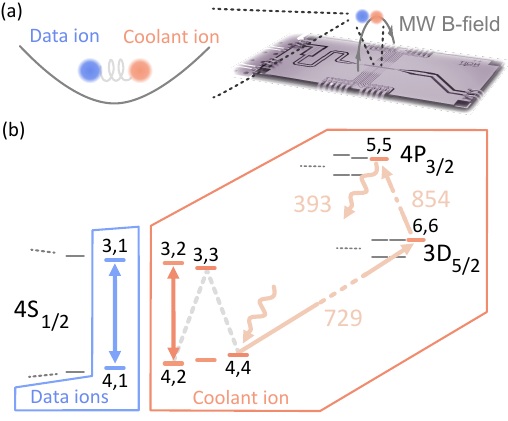}
    \caption{
    \textbf{Same-isotope sympathetic cooling scheme.}
    \textbf{(a)} Data ions and coolant ions are trapped within the same potential well, leading to shared motional modes.
    Coolant ions are used to cool the shared motion of all ions without affecting the information stored in the electronic states of the data ions.
    Errors in quantum logic operations between data ions, which make use of this shared motion, are thus reduced.
    A surface-electrode Paul trap is used to confine the two ions 40~$\upmu$m above the trap surface and an integrated microwave waveguide is used to generate a magnetic field for driving pulsed sideband cooling, individual ion addressing, and quantum logic operations (ions not to scale).
    \textbf{(b)} We make use of the Zeeman splittings between levels in the ground state manifold to isolate the data ions' microwave transition from a set of microwave and laser transitions used in the coolant ions.
    We use $^{43}$Ca$^{+}$ ions for which states $\ket{F = f, M = m}$ in the energy diagram are labelled as $f,m$.
    In each cooling cycle, we drive the motional sideband of the transition $\ket{4,2}$ to $\ket{3,2}$ with microwaves.
    Laser pulses at $729$~nm and $854$~nm, and microwave $\pi$-pulses (grey dashed lines) are then used to re-initialise the coolant ion before the next cycle.
    }
    \label{fig:fig1}
\end{figure}

\begin{figure}[t]
    \centering
    \includegraphics[width=0.45\textwidth]{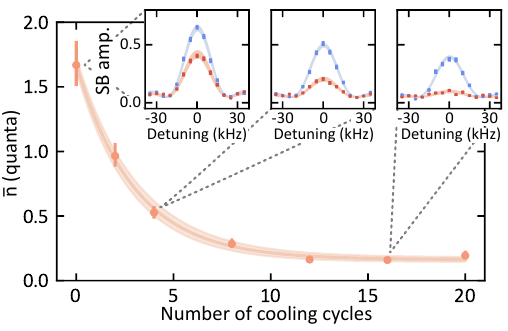}
    \caption{
    \textbf{Microwave-driven sideband cooling.}
    Measurements of the average phonon occupation $\bar{n}$ of the in-plane radial rocking mode are shown after applying a varying number of cooling cycles as described in Fig.~\ref{fig:fig1}.
    An exponential fit (shaded area) reveals a cooling rate of 0.27(1)~quanta/cycle at $\bar{n}=1$ and a minimum occupation of $\bar{n}=0.16(2)$ after 16 cooling cycles.
    The mode occupation is determined through sideband thermometry, implemented with a Raman laser interrogating the sidebands of the $\ket{4, 4}$ to $\ket{4, 3}$ transition.
    In the insets, we show the blue and red sideband amplitudes for 0, 4 and 16 cooling cycles.
    We perform detuning scans of both sidebands which are jointly fitted (shaded areas) to obtain $\bar{n}$.
    }
    \label{fig:fig2}
\end{figure}
%

%%%%%%%%%%%%%% "OVERVIEW"
%
In this Letter, we present a method for sympathetic cooling which uses ions of the same species and isotope, as well as data qubits stored in atomic clock states within the ground state manifold.
We make use of the large Zeeman splitting, created by a homogeneous static magnetic field, to separate data and cooling transitions in frequency-space, which can then be addressed even with global fields.
This method thus does not require tightly-focused lasers or the mapping of quantum information to metastable states.
The sympathetic cooling is driven by global near-field microwaves (MWs) and relies on the same control techniques as used for MW-driven logic, therefore imposing little experimental overhead.
Using the scheme outlined in Fig.~\ref{fig:fig1}, we demonstrate near-ground-state cooling of the in-plane radial rocking mode to an average phonon occupation $\bar{n}$ of 0.16(2) quanta.
We interleave cooling cycles with single-qubit gate sequences on the data ion~\cite{knill2008,omalley2015,sepiol2019,epstein2014}, and measure an error of $1.7(4) \times 10^{-4}$ per cooling cycle on the data ion.
The scheme is demonstrated using $^{43}$Ca$^{+}$ ions, but is applicable to all ion species with metastable states and with $\geq$~4 states within the ground state manifold.
%

%%%%%%%%%%% TRAP AND STATES
%
Experiments are carried out at room temperature in a segmented surface-electrode Paul trap (presented in Ref.~\cite{weber2022}) used to confine two $^{43}$Ca$^{+}$ ions 40~$\upmu$m above the chip surface: one data ion and one coolant ion.
An on-chip MW resonator is used together with the MW drive chain detailed in~\citesec{SuppInfoLabel}{\ref{SI:mw_chain}} to generate a magnetic field for driving transitions within the ground manifold $4\text{S}_{1/2}$.
Throughout this manuscript, we will refer to hyperfine states within the $4\text{S}_{1/2}$ manifold with the notation $\ket{f, m} = \ket{4\text{S}_{1/2}, F = f, M = m}$.
Our data qubit is defined by the states $\ket{4, 1}$ and $\ket{3, 1}$ (see Fig.~\ref{fig:fig1}), which form a ``clock'' qubit at a static magnetic field strength of 28.8 mT and are connected by a magnetic dipole transition with a frequency of $3.123$~GHz.
Our coolant ion makes use of transitions between states with $M > 1$, chiefly the motional sideband of the transition $\ket{4, 2}$ to $\ket{3, 2}$ with a frequency splitting of $2.908$~GHz, used for resolved sideband cooling.
%

%%%%%%%%%%% STATE-PREP
%
The scheme starts by preparing the ions in different states within the $4\text{S}_{1/2}$ manifold: $\ket{3, 1}$ (data ion) and $\ket{3, 2}$ (coolant ion).
To do so, we begin by preparing both ions in the state $\ket{4, 4}$ through dark-resonance Doppler cooling~\cite{allcock2016} followed by MW-enhanced optical pumping~\cite{harty2014}.
We then transfer both ions' states to $\ket{3, 1}$ using MW $\pi$-pulses.
Next, the addressing scheme outlined in Ref.~\cite{leu2023} is used to transfer the data population to the state $\ket{4, 1}$ whilst the coolant population remains in the state $\ket{3, 1}$.
Finally, MW $\pi$-pulses are used to transfer the coolant population to $\ket{3, 2}$ and then the data population to $\ket{3, 1}$.
Further details on the state preparation can be found in~\citesec{SuppInfoLabel}{\ref{SI:scheme_overview}}.

\begin{figure*}[t]
    \centering
    \includegraphics[width=0.95\textwidth]{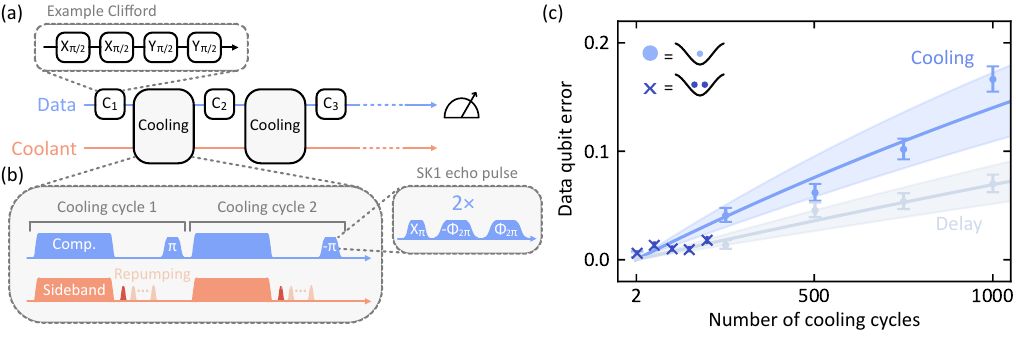}
    \caption{
    \textbf{Data qubit protection during sympathetic cooling.}
    \textbf{(a)} Interleaved randomised benchmarking (IRB) scheme used to measure the error induced on the data qubit during sympathetic cooling.
    Cooling cycles are interleaved between Clifford gates in a single-qubit randomised benchmarking sequence applied to the data qubit which, in the absence of errors, would return the qubit to its initial state.
    Deviations from this expected result reveal the error induced by the cooling cycles.
    \textbf{(b)} Data qubit protection scheme used during cooling cycles.
    During the microwave sideband pulse, a ``compensation'' pulse, detuned from the qubit transition, cancels the ac Zeeman shift induced on the data qubit.
    Cooling cycles are carried out in pairs, each ending with a MW $\pi$-pulse on the data qubit, implementing a spin-echo scheme to cancel residual ac Zeeman shifts.
    SK1 composite pulses are used to protect against pulse amplitude fluctuations from thermal transients ($\Phi = \arccos(-1/8)$).
    \textbf{(c)} Data qubit error measured through IRB.
    The majority of the data was acquired using a single ion (blue circles) to benefit from an increased trapping lifetime, and was validated with two ions for shorter gate sequences (dark blue crosses).
    An exponential fit yields an error of 1.7(4)$\times 10^{-4}$ per cooling cycle (blue shaded area).
    By replacing cooling cycles with delays of equal duration, we extract an ``idle error'' of 0.7(2)$\times 10^{-4}$ per cycle (grey).
    }
    \label{fig:fig3}
\end{figure*}

%%%%%%%%% MOTIONAL SIDEBAND + REPUMPING
%
After state preparation, pulsed resolved sideband cooling is used to cool the shared motion of the ions~\cite{diedrich1989,roos1999,king1998,monroe1995,marzoli1994}.
With the coolant ion initialised in $\ket{3, 2}$, we drive the red motional sideband of the transition $\ket{3, 2}\leftrightarrow\ket{4, 2}$ using microwaves~\cite{ospelkaus2011,khromova2012,weidt2015}.
After applying the 500~$\upmu$s sideband pulse, population transferred from $\ket{3, 2}$ to $\ket{4, 2}$ is associated with the removal of energy from the motional mode.
We thus refer to the population in $\ket{4, 2}$ as the ``cooled'' population, with the ``remaining'' population still in $\ket{3, 2}$.
Next, the cooled population is transferred to the state $\ket{4, 4}$ and the remaining population to the state $\ket{3, 3}$ using MW $\pi$-pulses.
To ``recombine'' the coolant ion populations, a $729$~nm $\pi$-pulse transfers the cooled population from $\ket{\text{S}_{1/2}, 4, 4}$ to $\ket{\text{D}_{5/2}, 6, 6}$, whereupon a $854$~nm pulse repumps the population back to $\ket{\text{S}_{1/2}, 4, 4}$ via $\ket{\text{P}_{3/2}, 5, 5}$.
In parallel, whilst the cooled population lies in the metastable state $\ket{\text{D}_{5/2}, 6, 6}$, we use a MW $\pi$-pulse to transfer the remaining population to $\ket{4, 4}$, such that the cooled and remaining populations are recombined after the repumping.
Imperfections in the $729$~nm $\pi$-pulse, and in the closed-loop formed by the repumping cycle, would leave population in $\ket{3, 3}$, which we reduce by swapping $\ket{4, 4}$ and $\ket{3, 3}$ populations using a MW $\pi$-pulse and repeating the ``recombination'' sequence (see~\citesec{SuppInfoLabel}{\ref{SI:scheme_overview}} for further details of the scheme and~\citesec{SuppInfoLabel}{\ref{SI:cooling:repumping}} for an assessment of the impact of imperfections in the repumping sequence).
%

%%%%%%%%%%%%%% IMPLEMENTATION
%
We demonstrate the scheme using the in-plane radial rocking mode of the two-ion crystal ($\omega=2\pi \times 5.37$~MHz), which is initially dark-resonance cooled to $1.7(2)$~quanta~\cite{allcock2016}, and heats up at a rate of $13(5)$~quanta/s.
The average phonon occupation $\bar n$ is determined for a varying number of cooling cycles through motional sideband asymmetry -- we scan the detuning of the red and blue motional sidebands of the $\ket{4, 4}$ to $\ket{4, 3}$ transition using a Raman laser setup and jointly fit the results, as shown in Fig.~\ref{fig:fig2}.
After 16 cooling cycles, we lower the occupation to $\bar{n}=0.16(2)$~quanta.
At $\bar{n}=1$~quanta, we determine a cooling rate of 0.27(1)~quanta/cycle, or 290(10)~quanta/s.
The cooling rate can be increased by reducing the 920~$\upmu$s cycle time, for example through cryogenic operation which has been shown to enable a $6 \times$ increase in ion-motion interaction strength~\cite{weber2024}.
Reliable sideband cooling requires mitigating changes in the motional sideband frequency occuring on different timescales.
The $\sim$~-30~kHz ac Zeeman shift on the $\ket{4, 2}$ to $\ket{3, 2}$ transition, induced by sideband driving, is found to change by $\sim$~5~kHz over the course of a 500~$\upmu$s pulse, which exceeds the motional sideband linewidth of $2.1(1)$~kHz.
We correct for this drift by simultaneously driving a second ``compensation'' MW tone, detuned by $-1.37$~MHz from the $\ket{4, 2}$ to $\ket{3, 2}$ transition, to apply an ac Zeeman shift in the opposite direction, see~\citesec{SuppInfoLabel}{\ref{SI:cooling:shifts}}.
This ensures that any amplitude drifts common to both sideband and compensation tones do not result in a net ac Zeeman shift on the sideband transition.
Even with this compensation pulse, the sideband frequency changes on a pulse-to-pulse basis, converging to a $\sim$~3~kHz shift after $\sim$~10 pulses, see~\citesec{SuppInfoLabel}{\ref{SI:cooling:shifts}}.
We mitigate this by applying a ``warmup'' sideband pulse immediately after state-preparation.
Despite these mitigation strategies, the cooling rate is still limited by $\sim$~250~Hz sideband frequency changes, captured through diagnostic measurements interleaved in between cooling measurements, see~\citesec{SuppInfoLabel}{\ref{SI:cooling:shifts}}.
All drifts mentioned above are at least partially due to thermal transients in the MW drive chain and/or on-chip MW resonator, induced by the $\sim$~1~W power of the sideband cooling drive.

%%%%%%%%%% DATA QUBIT PROTECTION SCHEME
%
As a result of the above scheme being applied to the coolant ion, the data ion experiences multiple frequency shifts which must be corrected to protect the quantum information stored in its qubit.
% See notes from 01/06 for simulation on pulse ramp up adiabaticity
Driving the sideband and compensation pulses around the $\ket{4, 2}$ to $\ket{3, 2}$ transition drives negligible population transfer on the data qubit transition $\ket{4, 1}$ to $\ket{3, 1}$ due to the large frequency detuning ($\sim$~200~MHz relative to a magnetic dipole Rabi frequency of $\sim$~1~MHz) and the use of a 400~ns pulse ramp-up/down time.
However, these pulses do induce an ac Zeeman shift of $\sim$~16.4~kHz on the data qubit transition.
We cancel this shift by simultaneously driving a second ``compensation'' MW tone, detuned by $+8.37$~MHz from the data qubit transition, to apply an ac Zeeman shift in the opposite direction.
Due to changes in the relative amplitudes of the different MW drive frequencies, induced by thermal drifts in the MW drive chain and/or on-chip MW resonator, there remains a fluctuating ac Zeeman shift of $\sim$~100~Hz.
This residual detuning, as well as the additional  $\sim$~5~kHz frequency shifts induced by the short MW transfer pulses used to re-initialise the coolant ion, are corrected by applying cooling cycles in groups of two, interleaved by spin-echo pulses on the data qubit transition~\cite{hahn1950}, see Fig.~\ref{fig:fig3}.
%

%%%%%%%%%% DATA QUBIT PROTECTION RESULTS
%
The resulting error induced on the data qubit is 1.7(4)$\times 10^{-4}$ per sympathetic cooling cycle, evaluated through interleaved randomised benchmarking (IRB)~\cite{knill2008,omalley2015,sepiol2019,epstein2014}.
In this IRB measurement, two cooling cycles are applied after every single-qubit Clifford gate in a randomised benchmarking sequence (see Fig.~\ref{fig:fig3}).
IRB is mostly carried out with a single data ion in the trap to benefit from an increased trapping lifetime and validated with two ions (one data, one coolant) for shorter gate sequences.
By replacing the cooling cycle with a delay of equal duration we extract an idle error of $0.7(2)\times 10^{-4}$ per cooling cycle.
To quantify the different sources of error, we repeat this measurement for each subsection of the cooling cycle, see~\citesec{SuppInfoLabel}{\ref{SI:data_qubit}}.
The MW sideband pulse, including associated ac Zeeman shift compensation pulses, induces the largest error (1.2(3)$\times 10^{-4}$ per cooling cycle), followed by the MW transfer pulses (0.3(1)$\times 10^{-4}$ per cooling cycle).
The error induced by laser pulses was indistinguishable from the idle error.
%

%%%%%%%%%% DATA QUBIT PROTECTION DISCUSSION
%
The error on the data qubit due to sympathetic cooling appears to be technical in nature, with the majority of the errors attributed to thermal drifts in the MW drive chain.
Thermal drifts can lead to residual ac Zeeman drifts which are not constant over the course of a pair of cooling cycles, and therefore are not corrected by the spin-echo pulses.
These drifts also lead to changes in the amplitudes of both the spin-echo MW pulses and the pulses used in implementing the single-qubit Clifford gates.
We use SK1 composite pulses to mitigate these errors~\cite{kabytayev2014,brown2004}.
In this demonstration, thermal effects are exacerbated as we are operating the trap at room temperature, leading to higher on-chip resistivity compared to cryogenic temperatures~\cite{weber2022}, combined with a decreased thermal conductivity of the chip's sapphire substrate.
Cryogenic operation would therefore be a first route towards improving the error per cycle.
A multi-frequency implementation of MW power management used in Ref.~\cite{weber2024} would be another route for improvement.
Errors may also be reduced by using a sideband cooling transition further separated in frequency from the data qubit to reduce ac Zeeman shifts.
One could make use of a different transition within the $^{43}$Ca$^{+}$ $4\text{S}_{1/2}$ manifold, or utilise a different ion species and/or static magnetic field to obtain a larger Zeeman splitting.
Sideband cooling may also be implemented using the motional sideband of the 729~nm transition; however this requires a technically-demanding high-power, stable quadrupole laser, see~\citesec{SuppInfoLabel}{\ref{SI:cooling_transition}}.
Ultimately, errors on the data qubit will be limited by scattering of the $393$~nm photon emitted by the coolant ion during the repumping step, which we estimate to induce an error of $1.6(9)\times 10^{-6}$ per cooling cycle, see~\citesec{SuppInfoLabel}{\ref{SI:data_qubit}}.
%

%%%%%%%%%% CONCLUSION
%
In conclusion, we have demonstrated a single-isotope sympathetic cooling method where both logical and cooling operations lie in the ground state manifold and are driven with microwaves.
Using this scheme, we cool the ions' shared motion whilst inducing an error of 1.7(4)$\times 10^{-4}$ per cooling cycle on a data qubit, an additional error of 1.0(4)$\times 10^{-4}$ per cycle over the qubit's idle error.
With the radial rocking mode used here (heating rate $=$ 13(5)~quanta/s, cooling rate $=$ 0.27(1)~quanta/cycle or 290(10)~quanta/s), maintaining $\bar{n}=1$ quanta would require $\sim$~50 cooling cycles per second.
Maintaining $\bar{n}=1$ quanta is sufficient to obtain the lowest entangling gate errors of any quantum computing platform~\cite{hughes2025} using electronically-driven logic.
The contribution of our scheme to the error per two-qubit gate would therefore be $\sim$~$10^{-6}$ per gate, much lower than the total error of state-of-the-art entangling gates~\cite{hughes2025}.
Furthermore, there are clear paths to further lowering the technically dominated error rate, as well as increasing the cooling rate, such as using cryogenic operation and microwave power management, with a fundamental limit in error two orders of magnitude lower than the measured error.
We note that the ability to preserve quantum coherence during same-isotope sympathetic cooling has recently been demonstrated at Duke University~\cite{ranawat2026}.
This method used a tightly-focused laser beam to map the coolant ion population to a state further separated in frequency from the data qubit whereupon a global beam was used to drive the motional sideband of the quadrupole transition.

\vspace{5mm}
\textbf{Acknowledgments}
We thank B. Nichols and R. Srinivas for the 729~nm laser system used in this work.
This work was supported by the U.S. Army Research Office (ref. W911NF-24-1-0380) and the U.K. EPSRC Quantum Computing and Simulation Hub (ref. EP/Z53318X/1).
M.C.S. acknowledges support from Balliol College, Oxford.
A.D.L. acknowledges support from Oxford Ionics Ltd.
K.M. acknowledges support from the Japan Science and Technology Agency (JST) ASPIRE Japan (ref. JPMJAP2319), and the JST Moonshot Research and Development program (ref. JPMJMS2063).

\vspace{5mm}
\textbf{Author Contributions}
M.C.S., E.V., A.D.L., N.D., A.A., K.M. and M.F.G. developed and maintained the experimental apparatus.
M.C.S. and E.V. obtained the data and carried out the data analysis.
M.C.S. and E.V. wrote the manuscript with contributions from all authors.
M.F.G. and D.M.L. supervised the work.

\bibliography{library}

\FloatBarrier
\clearpage
\onecolumngrid
\begin{center}
{\Large \textbf{Supplementary information}}
\end{center}
\makeatletter
   \renewcommand\l@section{\@dottedtocline{2}{1.5em}{2em}}
   \renewcommand\l@subsection{\@dottedtocline{2}{3.5em}{2em}}
   \renewcommand\l@subsubsection{\@dottedtocline{2}{5.5em}{2em}}
\makeatother
\let\addcontentsline\oldaddcontentsline% Restore \addcontentsline

\renewcommand{\thesection}{\arabic{section}}

% \twocolumngrid

\let\oldaddcontentsline\addcontentsline% Store \addcontentsline
\renewcommand{\addcontentsline}[3]{}% Make \addcontentsline a no-op
\let\addcontentsline\oldaddcontentsline% Restore \addcontentsline
\renewcommand{\theequation}{S\arabic{equation}}
\renewcommand{\thefigure}{S\arabic{figure}}
\renewcommand{\thetable}{S\arabic{table}}
\renewcommand{\thesection}{S\arabic{section}}
\setcounter{figure}{0}
\setcounter{equation}{0}
\setcounter{section}{0}

\newcolumntype{C}[1]{>{\centering\arraybackslash}p{#1}}
\newcolumntype{L}[1]{>{\raggedright\arraybackslash}p{#1}}

% \tableofcontents
\FloatBarrier

%!TEX root = output/arxiv.tex

The sections of the supplementary information each provide details on a figure of the main text.
In section~\ref{SI:scheme_overview}, we provide a more detailed overview of the sympathetic cooling scheme shown in Fig.~\ref{fig:fig1}, as well as our approach to state-preparation and measurement in our implementation of it.
In section~\ref{SI:cooling}, we provide additional data and simulations contributing to the characterisation and calibration of the sideband cooling measurements of Fig.~\ref{fig:fig2}.
In section~\ref{SI:data_qubit}, we present measurements used to characterise the protection of the data qubit shown in Fig.~\ref{fig:fig3}.
Lastly, in section~\ref{SI:mw_chain}, we provide a schematic of the microwave drive chain used to generate the microwave pulses used in this work.

\section{Experimental scheme}\label{SI:scheme_overview}
In this section, we provide a more detailed overview of the scheme and our implementation of it, complementing the information shown in Fig.~\ref{fig:fig1}.
We first cover the state-preparation and measurement techniques used before and after the sympathetic cooling and data qubit manipulations.
We then provide a pulse-by-pulse description of the sideband pulse delivery and coolant re-initialisation process.
Throughout this manuscript, we will refer to hyperfine states within the $4\text{S}_{1/2}$ manifold with the notation $\ket{f, m} = \ket{4\text{S}_{1/2}, F = f, M = m}$.

\subsection{State-preparation and measurement}
\begin{figure}[b]
    \centering
    \includegraphics[width=0.9\textwidth]{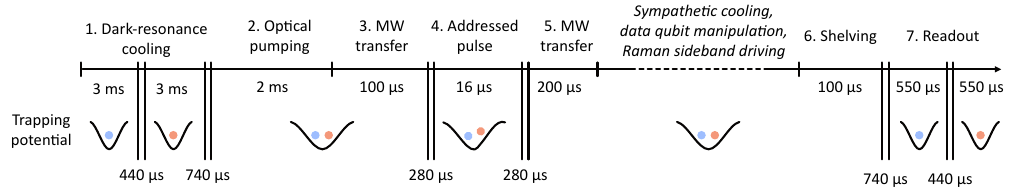}
    \caption{
    \textbf{Overview of state-preparation and measurement sequence.}
    The individual steps follow the numbering used in the written description above and schematically show the use of single-ion wells, or twisted/untwisted two-ion wells with data and coolant ions shown in blue and orange, respectively.
    Timings below the figure refer to the time required to update the trapping potential well.
    }
    \label{fig:spam}
\end{figure}
The ions must first be prepared into different states within the ground manifold $4\text{S}_{1/2}$: $\ket{3, 1}$ (data ion) and $\ket{3, 2}$ (coolant ion).
The scheme used is as follows:
\begin{enumerate}
    \item Both ions are subjected to dark-resonance Doppler cooling~\cite{allcock2016}.
    Since the two-ion lifetime in this trap is limited to $\sim$~2 minutes, we keep data and coolant ions in separate wells during this process, shuttling the ions to the centre of the trap to be laser-cooled in turn.
    All motional modes are cooled in this process, leading to a radial rocking mode occupation of $\bar n=1.7(2)$ after merging the two ion wells and performing the state preparation routine described below.
    % 2
    \item Electronic state preparation is carried out after merging the two single-ion wells. Both ions are first prepared in the state $\ket{4, 4}$ through microwave-enhanced optical pumping~\cite{harty2014}.
    % 3
    \item The electronic states of both ions are then transferred to $\ket{3, 1}$ using MW transfer pulses acting identically on both ions.
    % 4
    \item The addressing scheme outlined in Ref.~\cite{leu2023} is subsequently used to transfer the data ion to the state $\ket{4, 1}$ whilst the coolant ion remains in the state $\ket{3, 1}$.
    Here, DC voltages applied to on-chip electrodes rotate the ion crystal within the near-field microwave amplitude gradient such that both ions are driven by different Rabi frequencies, enabling single-ion addressing through composite pulse sequences.
    % 5
    \item Finally, MW transfer pulses are used to transfer the coolant ion to $\ket{3, 2}$ and the data ion to $\ket{3, 1}$, which ends the state-preparation sequence. At this stage, sympathetic cooling combined with either data qubit manipulations or Raman sideband thermometry are carried out.
    % 6
    \item The readout sequence first consists of MW transfer pulses to transfer a desired state in either the data qubit (for interleaved benchmarking measurements), or the coolant ion (for thermometry), to the state $\ket{4, 4}$.
    The population in $\ket{4, 4}$ is then shelved to the metastable $5\text{D}_{5/2}$ level using 393~nm and 850~nm laser pulses~\cite{myerson2008}.
    % 7
    \item At this stage the ions are split into two wells, and are subjected one after the other to the Doppler cooling beams, whereupon the state of each ion is inferred from the fluorescence level.
    Lastly, either the process starts anew for further data acquisition, or the ions are kept in separate wells and cooled intermittently whilst idling.
\end{enumerate}
The resulting state-preparation and measurement (SPAM) errors are $5(1) \times 10^{-3}$ (data ion) and $3(1) \times 10^{-3}$ (coolant ion), dominated by the errors from the MW transfer pulses.
The process is shown schematically in Fig.~\ref{fig:spam}.

\subsection{Sideband driving scheme}
\begin{figure}[b]
    \centering
    \includegraphics[width=0.9\textwidth]{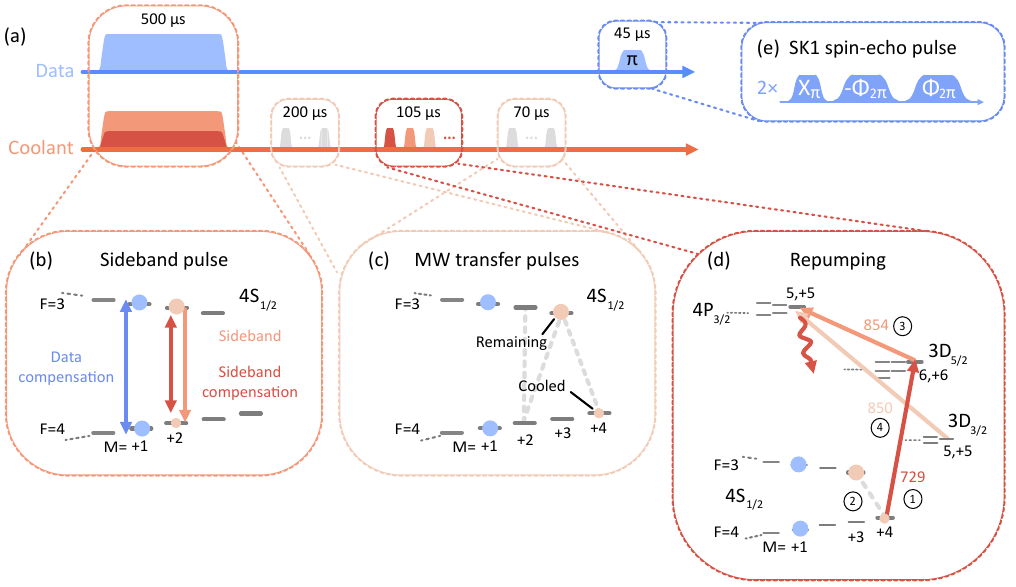}
    \caption{
    \textbf{Overview of the sympathetic cooling scheme.}
    \textbf{(a)} A timeline describing the pulses applied during one cycle of sympathetic cooling is shown for both data (blue) and coolant (orange) ions.
    \textbf{(b)} Whilst the motional sideband is driven on the coolant transition, two detuned compensation pulses are applied to correct for ac Zeeman shifts induced on both the coolant and data transitions.
    \textbf{(c)} MW transfer pulses (grey dashed lines) are used to transfer the coolant ion population around the ground state $4\text{S}_{1/2}$ manifold. Here the population distribution is shown after the sideband pulse has been applied and with the coolant states transferred to their final position before repumping.
    \textbf{(d)} The coolant ion population is repumped through a series of laser and MW pulses.
    \textbf{(e)} Finally, a spin-echo pulse is applied on the data transition to correct for frequency shifts induced by the cooling sequence.
    }
    \label{fig:scheme_overview}
\end{figure}

To cool the ions' motion, we drive the motional sideband on the transition $\ket{4, 2}$ to $\ket{3, 2}$.
To drive this sideband whilst compensating for ac Zeeman shift drifts, we simultaneously apply the following microwave tones (see Fig.~\ref{fig:scheme_overview}(b)):
\begin{itemize}
    \item The sideband tone (2.908~GHz + 5.37~MHz), which drives the motional sideband on the transition $\ket{4, 2}$ to $\ket{3, 2}$ (the in-plane radial rocking mode frequency is 5.37~MHz).
    % Compensation detuning = +4 MHz from the RSB
    \item The sideband compensation tone (2.908~GHz - 1.37~MHz), which corrects for fluctuations in the ac Zeeman shift induced by the sideband tone on the coolant transition $\ket{4, 2}$ to $\ket{3, 2}$ -- see Sec.~\ref{SI:cooling} for more details.
    % Cancellation detuning = +3 MHz from the BSB
    \item The data compensation tone (3.123~GHz + 8.37~MHz), which corrects for ac Zeeman shift fluctuations induced by the two tones above on the data transition $\ket{4, 1}$ to $\ket{3, 1}$ -- see Sec.~\ref{SI:data_qubit} for more details.
\end{itemize}
Any other zeeman shifts on the data qubit are corrected for by applying a spin-echo pulse on the data qubit between cooling cycles, as shown in Fig.~\ref{fig:scheme_overview}(e), with more detail in Sec.~\ref{SI:data_qubit}.
A schematic of the experimental setup used to generate these tones is shown in Fig.~\ref{fig:mw_chain}.

\subsection{Re-initialisation of the coolant ion}

After applying the sideband pulse, population transferred from $\ket{3, 2}$ to $\ket{4, 2}$ is associated with the removal of energy from the motional mode.
We thus refer to the population in $\ket{4, 2}$ as the ``cooled'' population, with the ``remaining'' population still in $\ket{3, 2}$.
To prepare the coolant for another round of cooling, we first transfer the ``cooled'' population to the state $\ket{4, 4}$ and the ``remaining'' population to the state $\ket{3, 3}$ using MW transfer pulses (see Fig.~\ref{fig:scheme_overview}(c)).
The scheme to recombine the ``cooled'' and ``remaining'' populations then carries on as described below and in Fig.~\ref{fig:scheme_overview}(d):
\begin{enumerate}
    % 1
    \item A $729$~nm $\pi$-pulse (13~$\upmu$s) is driven on the transition $\ket{4\text{S}_{1/2}, 4, 4}$ to $\ket{3\text{D}_{5/2}, 6, 6}$.
    % 2
    \item A MW $\pi$-pulse is applied to transfer the ``remaining'' population from $\ket{3, 3}$ to $\ket{4, 4}$.
    % 3
    \item An $854$~nm pulse (2~$\upmu$s) is driven on the transition $\ket{3\text{D}_{5/2}, 6, 6}$ to $\ket{4\text{P}_{3/2}, 5, 5}$. The population in the state $\ket{4\text{P}_{3/2}, 5, 5}$ will (mostly) decay back to the state $\ket{4\text{S}_{1/2}, 4, 4}$, emitting a $393$~nm photon.
    However, during this process, some of the population in the state $\ket{4\text{P}_{3/2}, 5, 5}$ may decay to other states, exiting the closed loop formed by the repumping cycle, see Sec.~\ref{SI:cooling}.
    Additionally, imperfections in the $729$~nm $\pi$-pulse will leave population in the state $\ket{4\text{S}_{1/2}, 3, 3}$ at this stage.
    To recombine the population in the state $\ket{4\text{S}_{1/2}, 3, 3}$, we apply another MW $\pi$-pulse on the transition $\ket{4\text{S}_{1/2}, 3, 3}$ to $\ket{4\text{S}_{1/2}, 4, 4}$ and repeat steps \textbf{1} to \textbf{3}.
    \item Lastly, an $850$~nm pulse (2~$\upmu$s) is applied on the transition $\ket{3\text{D}_{3/2}, 5, 5}$ to $\ket{4\text{P}_{3/2}, 5, 5}$ to empty population which has become trapped in the $\text{D}_{3/2}$ manifold.
\end{enumerate}

\section{Microwave-driven sideband cooling}~\label{SI:cooling}

In this section, we give further details on the microwave-driven resolved sideband cooling measurements presented in Fig.~\ref{fig:fig2}.

\subsection{Mitigating frequency shift errors}~\label{SI:cooling:shifts}

\begin{figure}[h!]
    \centering
    \includegraphics[width=0.9\textwidth]{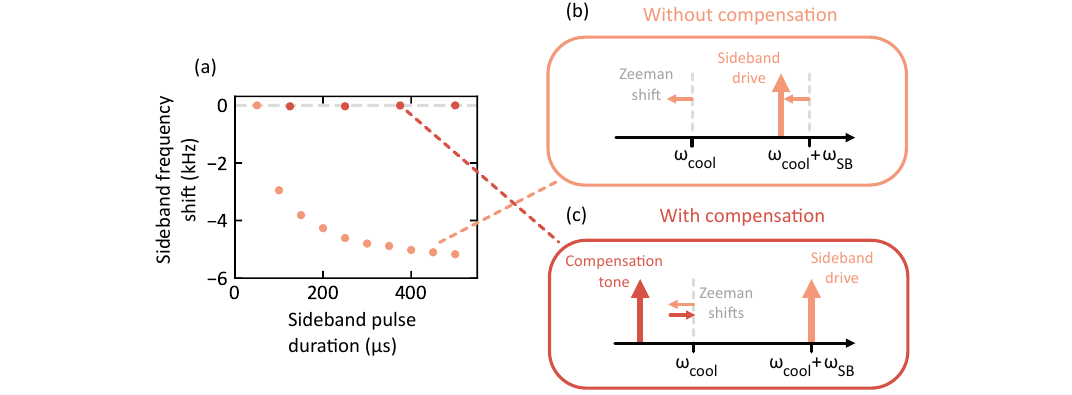}
    \caption{
    \textbf{Intra-pulse frequency shift and compensation}
    \textbf{(a)} Measurements of frequency shifts on the coolant transition ($\ket{4, 2}$ to $\ket{3, 2}$), extracted from Ramsey measurements with a varied sideband pulse duration applied.
    We distinguish cases where the sideband tone alone was applied (orange dots) or the sideband and compensation tones together (red) which impose an ac Zeeman shift of $\sim$~5~kHz or no shift respectively.
    \textbf{(b)} Sideband driving scheme without compensation.
    The sideband tone is detuned from the transition $\ket{4, 2}$ to $\ket{3, 2}$ with frequency $\omega_\text{cool}$ by the radial mode frequency $\omega_\text{SB}$.
    Off-resonant driving of the $\ket{4, 2}$ to $\ket{3, 2}$ transition induces an ac Zeeman shift.
    \textbf{(c)} Sideband driving scheme with compensation.
    A ``compensation tone'', with opposite detuning to the sideband drive, imposes a Zeeman shift of equal magnitude but opposite sign.
    This results in a nulled Zeeman shift which is insensitive to microwave amplitude changes common to both the sideband and compensation drive chains.
        }
    \label{fig:compensation}
\end{figure}

\textbf{Intra-pulse frequency shift and compensation.}
The MW sideband pulse induces an ac Zeeman shift of $\sim$~30~kHz on the coolant transition $\ket{4, 2}$ to $\ket{3, 2}$.
If this shift were constant, one could detune the frequency of the sideband pulse in order to resonantly drive the sideband.
However, when measuring the magnitude of this shift throughout the duration of the pulse, we observe a $\sim$~5~kHz change, see Fig.~\ref{fig:compensation}.
This is likely due to thermal drifts in the MW drive chain and/or on-chip MW resonator, induced by the $\sim$~1~W power of this drive, which result in fluctuations of the pulse amplitude.
Since this drift is significant compared to the sideband transition linewidth of 2.1(1)~kHz, it will lower the cooling efficiency, see Fig.~\ref{fig:mode_detuning}.
To correct for this shift, we simultaneously apply an additional ``sideband compensation'' MW tone detuned by -1.37~MHz from the $\ket{4, 2}$ to $\ket{3, 2}$ transition with an amplitude calibrated such that the total ac Zeeman shift vanishes.
Both pulses are generated by the same MW drive (see Fig.~\ref{fig:mw_chain}), so that thermal drifts and technical noise affect the amplitudes of both tones similarly and the corresponding ac Zeeman fluctuations mutually cancel.
This enables us to resonantly drive the motional sideband throughout a single instance of the 500~$\upmu$s sideband pulse, see Fig.~\ref{fig:compensation}.

\begin{figure}[h!]
    \centering
    \includegraphics[width=0.9\textwidth]{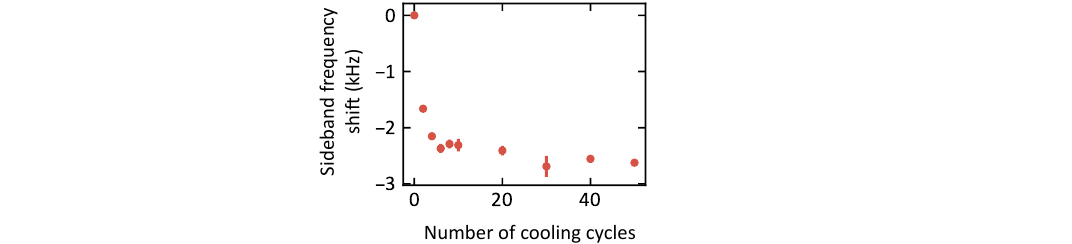}
    \caption{
    \textbf{Pulse-to-pulse frequency shift and compensation.}
    Here we show measurements of the sideband frequency shift following a varying number of cooling cycles.
    Shifts are extracted from detuning scans of the motional sideband.
    We observe n reproducible frequency shift which tends to saturate, motivating the use of ``warmup'' sideband pulse for 5~ms immediately after state-preparation to thermalise the MW drive chain.
        }
    \label{fig:warmup}
\end{figure}

\begin{figure}[h!]
    \centering
    \includegraphics[width=0.9\textwidth]{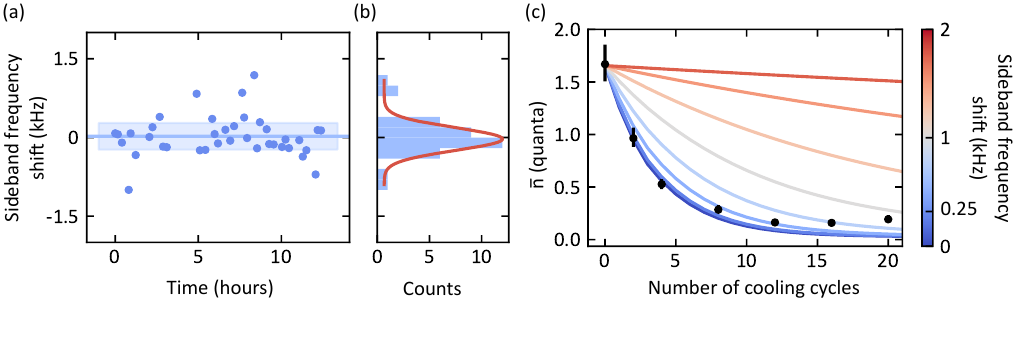}
    \caption{
    \textbf{Fluctuations in mode detuning.}
    \textbf{(a)} Calibrations of the in-plane radial rocking mode frequency, which were interleaved with the thermometry measurements shown in Fig.~\ref{fig:fig2}.
    Each calibration is determined by fitting a detuning scan of the $\ket{4, 4}$ to $\ket{4, 3}$ motional sideband, probed with our Raman laser setup.
    \textbf{(b)} Histogram of the change in mode frequency (blue), fitted with a normal distribution (red) to determine a half-width-half-maximum of 250(27)~Hz.
    \textbf{(c)} Simulated sideband cooling performance in the presence of a MW sideband detuning.
    The cooling measurements shown in Fig.~\ref{fig:fig2} are shown here again as black dots.
    The long-timescale frequency drifts shown in panels (a) and (b) thus lead to only a small reduction in cooling performance.
    However, these simulations motivate the need to compensate for fluctuations on shorter timescales, characterised in Figs.~\ref{fig:compensation} and \ref{fig:warmup}, which, if left uncorrected, would result in detuning errors of $\sim$~kHz.
    }
    \label{fig:mode_detuning}
\end{figure}

\vspace{10pt}
\textbf{Pulse-to-pulse frequency shift and warmup pulses.}
Despite the compensation scheme described in Fig.~\ref{fig:compensation}, when applying multiple repeats of the sympathetic cooling sequence there remains a pulse-to-pulse variation in the sideband frequency of $\sim$~3~kHz, see Fig.~\ref{fig:warmup}, which would again lead to inefficient cooling.
To eliminate these residual shifts, we apply a ``warmup'' sideband pulse to thermalise the MW drive chain.
This pulse is applied between steps 4 and 5 of the state-preparation and measurement sequence described in Sec.~\ref{SI:scheme_overview}.

\vspace{10pt}
\textbf{Longer term frequency shifts.}
Despite the ac Zeeman shift compensation scheme and warmup pulses, frequency shifts are still present when probed on a longer timescale.
We characterise these through measurements of the mode frequency interleaved with the cooling measurements shown in Fig.~\ref{fig:fig2}.
We find that the motional mode frequency varies by $\sim$~250~Hz from one measurement to the next (performed at approximately 15 minutes intervals), see Fig.~\ref{fig:mode_detuning}.
For this reason, we use a sideband pulse duration which is shorter than a $\pi$-pulse duration, moving away from the theoretically optimum sideband cooling pulse, but gaining in robustness to detuning errors.
The simulated impact of these detuning errors on the cooling rate is thus low, as shown in Fig.~\ref{fig:mode_detuning}(c).
However, reducing frequency drifts would enable a more optimum pulse time and an increased cooling rate.

\FloatBarrier

\subsection{The impact of imperfect coolant re-initialisation}~\label{SI:cooling:repumping}

\begin{figure}[t]
    \centering
    \includegraphics[width=0.9\textwidth]{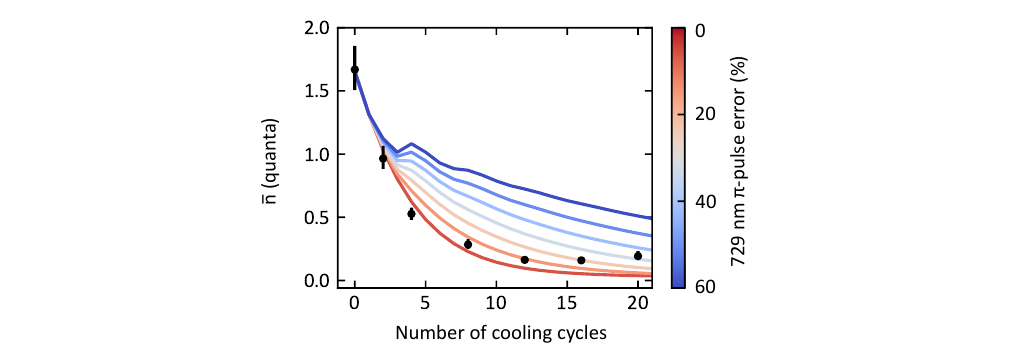}
    \caption{
    \textbf{Imperfect 729~nm $\pi$-pulse during repumping.}
    Simulated impact of the 729~nm $\pi$-pulse fidelity on the cooling efficiency.
    Here we do not correct for the 729~nm pulse error by performing multiple repeats of the repumping scheme.
    This is compared to the cooling measurements shown in Fig.~\ref{fig:fig2} (black), for which we perform two repeats of the repumping step, and during which the maximum 729~nm pulse error is 7~$\%$ (after 20 cooling cycles).
    }
    \label{fig:repumping_729_error}
\end{figure}
\begin{figure}[htbp]
    \centering
    \includegraphics[width=0.9\textwidth]{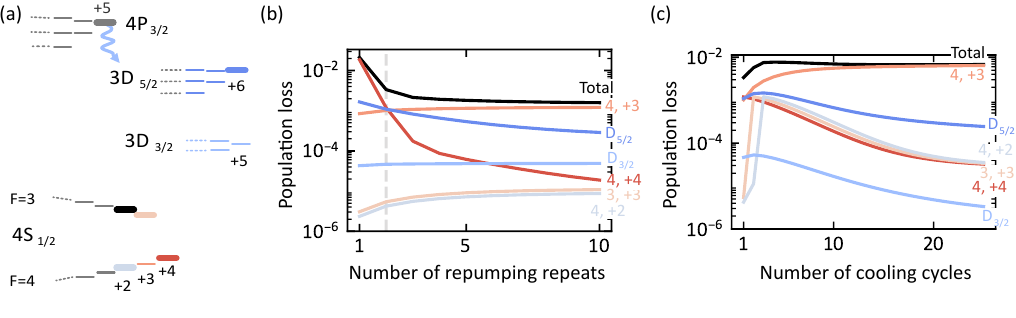}
    \caption{
    \textbf{Imperfect closed loop during repumping.}
    \textbf{(a)} States involved in the repumping process.
    We attempt to keep the population within the states highlighted in bold: the coolant transition states $\ket{3, 2}$ and $\ket{4, 2}$, and the states used for repumping $\ket{4, 4}$, $\ket{3, 3}$, $\ket{3\text{D}_{5/2}, 6, 6}$, and $\ket{4\text{P}_{3/2}, 5, 5}$.
    However, population may also decay back to six states within the $3\text{D}_{5/2}$ manifold (dark blue) (6~$\%$)-- or three states within the $3\text{D}_{3/2}$ manifold (light blue) (1~$\%$) -- which may subsequently be repumped by the $854$~nm and $850$~nm pulses to other states within the $4\text{P}_{3/2}$ manifold, exiting the repumping cycle.
    \textbf{(b)} Population exiting the repumping cycle for varying number of repeats of the repumping step within a single cooling cycle.
    The black line shows the total population lost, more precisely the population which has not returned to the desired state $\ket{3, 2}$ after repumping.
    Coloured lines show the population in other states, revealing that the majority of the lost population will be left in the state $\ket{4, 3}$.
    Simulations are carried out for a 729~nm pulse error of 5~$\%$ (corresponding to our 729~nm pulse error on the 12$^{\text{th}}$ cooling cycle)
    After two repeats of the repumping step, the error is reduced to $\sim 10^{-3}$.
    %.
    Population loss is thus reduced by repeating the repumping step twice (indicated by the grey dashed line), as described in Sec.~\ref{SI:scheme_overview}.
    \textbf{(c)} Population exiting the repumping cycle for varying number of repeats of the cooling cycle.
    Similarly to panel (b), we show simulation results for a 729~nm pulse error of 5~$\%$, with the same colour scheme, and for two repeats of the repumping step.
    We see that some of the population which has previously exited the closed loop can re-enter the cooling cycle via the MW swap pulses, as indicated by the total lost population (black) decreasing after $\sim$~4 cooling cycles.
    After 20 cooling cycles, the fraction of the population which has not returned to the desired state is 7 $\times 10^{-3}$.
    }
    \label{fig:repumping}
\end{figure}
The repumping scheme used in this work, described in Sec.~\ref{SI:scheme_overview}, does not form a perfect closed loop, meaning that some population is lost at every cooling cycle.
After the population is pumped to the short-lived state $\ket{4\text{P}_{3/2}, 5, 5}$, there are multiple channels through which it may decay.
The majority of the population (93~$\%$) will decay to the state $\ket{4, 4}$, remaining in the closed loop.
However, some population will decay back to the $3\text{D}_{5/2}$ (6~$\%$) and $3\text{D}_{3/2}$ manifolds (1~$\%$), see Fig.~\ref{fig:repumping}.
The $854$~nm pulse will repump the population which has decayed to the $3\text{D}_{5/2}$ back to the $4\text{P}_{3/2}$ manifold.

To repump the population lost in the $3\text{D}_{3/2}$ manifold, we add an $850$~nm pulse to the end of the repumping scheme.
However, from the $3\text{D}_{5/2}$ and $3\text{D}_{3/2}$ manifolds, the $854$~nm/$850$~nm pulses may transfer the population back to multiple states in the $4\text{P}_{3/2}$ manifold, which may not decay back to the state $\ket{4\text{S}_{1/2}, 4, 4}$, leading to a loss of population from the closed loop.
Additionally, the 729~nm pulse used to transfer population from the state $\ket{4, 4}$ to the state $\ket{3\text{D}_{5/2}, 6, 6}$ will not be a perfect $\pi$-pulse.
Due to the projection of the 729~nm beam onto the axial motional mode (heating rate 1600(200)~quanta/s), the fidelity of this $\pi$-pulse will be reduced to $\sim$~93~$\%$ after 20 cooling cycles, leading to an additional channel for loss of population from the closed loop.
Simulations of the impact of this error on the cooling rate can be found in Fig.~\ref{fig:repumping_729_error}.
This error is reduced by applying an additional MW $\pi$-pulse on the transition $\ket{3, 3}$ to $\ket{4, 4}$ and repeating the repumping scheme, as described in Sec.~\ref{SI:scheme_overview}, to repump the population left behind by the imperfect 729~nm $\pi$-pulse, see Fig.~\ref{fig:repumping}.
Due to the nature of this error (heating of the axial mode), this error will become more significant when this cooling scheme is interleaved with logical gates, during which the axial mode will continue to heat.
However, this can be corrected by applying further repeats of the repumping scheme, or choosing a 729~nm beam geometry with no projection onto the axial mode.
We estimate that the fraction of population exiting the repumping loop after 20 cooling cycles is 7 $\times 10^{-3}$, see Fig.~\ref{fig:repumping}(b).
This was simulated using the Atomic Physics project~\cite{atomicphysics}.
We determine that the majority of the lost population will be left in the state $\ket{4, 3}$.
One could bring this population back into the repumping cycle by transferring it to $\ket{4, 4}$ before carrying out an additional repumping cycle.
Alternatively, one could add polarisation control to the $854$~nm and $850$~nm pulses to reduce the population repumped to other states within the $4\text{P}_{3/2}$ manifold.
Another option would be to shuttle in a new coolant ion, that has been state-prepared in a different zone of the trap, once the population within the closed loop has been depleted.
In this work, we do not implement any of these options, since the fraction of population lost from the closed loop saturates with repetitions of the cooling process (Fig.~\ref{fig:repumping}(c)), and the extent of this loss is small enough to not significantly impact the cooling efficiency.
However, this is because we are performing multiple cycles of cooling without allowing mode heating to occur, and we thus have less ``cooled'' population to repump at every round of cooling.
If this cooling scheme were to be interleaved with logical gates to maintain $\bar{n} \approx 1$, the ``cooled'' population will not decrease, resulting in $\sim$~3 $\times 10^{-3}$ fraction of the population lost each time the sympathetic cooling sequence is applied.

\subsection{Choice of sideband cooling transition}\label{SI:cooling_transition}

In principle, any $\pi$-polarised transition within the ground manifold $4\text{S}_{1/2}$ could be used as the coolant transition for the sympathetic cooling scheme described in this work (since, for our system, the MW gradient is maximised for $\pi$-polarised transitions).
Using a coolant transition further separated in frequency-space from the data qubit transition (for example $\ket{4, -3}$ to $\ket{3, -3}$) would result in smaller ac Zeeman shifts on the data qubit, and therefore lower errors.
However, in this experimental setup, technical limitations result in lower cooling efficiencies for transitions further away from the data qubit.
Firstly, the on-chip MW resonator is centred around the data qubit frequency, meaning higher MW powers are required to drive transitions which are further detuned.
Secondly, the spatial profile of the MW field is frequency dependent.
The MW field gradient, used to couple the ion's spin and motion and therefore drive the sideband transition, decreases as the detuning from the data qubit transition is increased -- likely due to coupling between the MW and RF electrodes, see Ref.~\cite{weber2022} -- which results in a weaker spin-motion coupling.
Lastly, transitions further detuned from the data qubit are more susceptible to magnetic field fluctuations, resulting in lower coherence times.
Due to the $\sim$~500~$\upmu$s pulse times required to drive the sideband, lower coherence times ($<$~1~ms) would have a significant impact on the sideband dynamics.
For these reasons, the $\ket{4, 2}$ to $\ket{3, 2}$ transition was selected for sideband cooling.

One could also use the sideband of the $729$~nm transition $\ket{4\text{S}_{1/2}, 4, 4}$ to $\ket{3\text{D}_{5/2}, 6, 6}$ for cooling, which would eliminate the technical issues caused by thermal drifts from the high power MW sideband pulses (for example, the results presented in Figs.~\ref{fig:compensation}, and \ref{fig:warmup}).
However, this requires a high-power, stable quadrupole laser, which can be expensive, difficult to maintain, and complex to integrate into on-chip waveguides.
As a general principle, it is desirable to use the same technique to drive both sideband cooling transitions and entangling gates in this scheme, to minimise such overhead.

\section{Data qubit protection}~\label{SI:data_qubit}

\begin{figure}[h]
    \centering
    \includegraphics[width=0.9\textwidth]{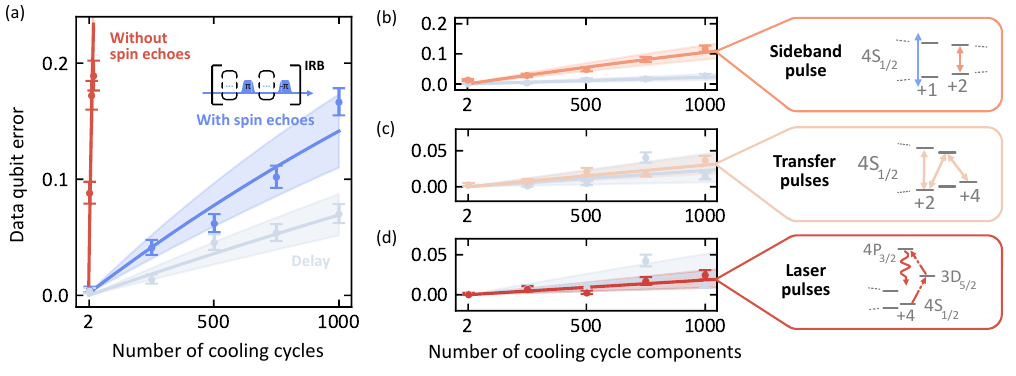}
    \caption{
    \textbf{Data qubit protection error budget.}
    \textbf{(a)} IRB measurements of the data qubit protection with spin-echo pulses (red) and without echo pulses (blue) are compared to IRB with a delay of equal duration to the cooling cycle (grey).
    The spin-echo pulses reduce the error induced on the data qubit from $1.5(5) \times 10^{-2}$ to $1.7(4) \times 10^{-4}$.
    Replacing the sympathetic cooling in the IRB sequence with delays of equal duration leads to a data qubit error of $7(2) \times 10^{-5}$ per delay (grey).
    \textbf{(b)} IRB measurements are then repeated but with only the sideband pulse part of the cooling cycle.
    The MW sideband pulse is the largest contribution to the total error on the data qubit with an error of $1.2(3) \times 10^{-4}$ per 500~$\upmu$s pulse (orange).
    $2(1) \times 10^{-5}$ of this error can be attributed to a delay of the same duration (grey).
    \textbf{(c)} IRB measurements with only the MW $\pi$-pulses required to transfer the coolant population around the ground state manifold.
    These induce ac Zeeman shifts of $\sim$~5~kHz on the data qubit, which would result in an error of $2.9(5) \times 10^{-3}$ per cooling cycle without any data qubit protection.
    Adding the spin-echo $\pi$-pulses reduces the error to $3(1) \times 10^{-5}$ per cooling cycle (pale orange).
    A delay of the combined duration of all MW $\pi$-pulses used in one cooling cycle causes an error of $2(2) \times 10^{-5}$ per cycle (grey).
    \textbf{(d)} The data qubit error of $2(1) \times 10^{-5}$ per cooling cycle induced by the laser pulses (red) used in the repumping scheme is indistinguishable from the decoherence that would occur on the qubit for delays of the same duration as the laser pulses (grey).
    }
    \label{fig:error_budget}
\end{figure}

In this section, we quantify the improvements in data qubit protection achieved through the use of Zeeman shift compensation and echo pulses.
All errors are characterised through interleaved randomised benchmarking (IRB) measurements, as presented for the full scheme in Fig.~\ref{fig:fig3}.
Fig.~\ref{fig:error_budget} gives an overview of the contributions to the total residual error on the data qubit from each step of the cooling scheme.
All IRB data presented here and in the main text is fitted to an exponential decay model~\cite{knill2008}, using a maximum likelihood estimation method, and uncertainties were determined using parametric bootstrapping, see Ref.~\cite{smith2025} for further details on the fitting.

\subsection{Error induced by the sideband pulse}
The sideband pulse (together with its compensation tone, see Sec.~\ref{SI:cooling}) induces an ac Zeeman shift of 16.48(2)~kHz on the data qubit.
A compensation tone, $+8.37$~MHz detuned from the data qubit transition, is applied in order to compensate for this shift.
This tone is generated by the same MW drive as the sideband and compensation tones (see Fig.~\ref{fig:mw_chain}), such that thermal drifts and technical noise affect the amplitudes of all tones similarly and the corresponding ac Zeeman fluctuations mutually cancel.
Due to a combination of thermal drifts and a limited calibration accuracy, a residual frequency shift on the order of $\sim$~100~Hz remains.
This shift is corrected by the spin-echo pulses following each cooling cycle, reducing the error induced by sideband driving to 1.2(3)$\times 10^{-4}$ per cooling cycle, see Fig.~\ref{fig:error_budget}(b).

\subsection{Error induced by the microwave $\pi$-pulses}~\label{SI:data_qubit_transfer}

Throughout the cooling cycle, population must be transferred between states of the ground state manifold using MW $\pi$-pulses.
These pulses induce ac Zeeman shifts between 3.4~kHz and 5~kHz on the data qubit (see Tab.~\ref{tab:transfer_pulses}).
Applying the spin-echo pulses after each cooling cycle reduces the error due to the transfer pulses to 0.3(1)$ \times 10^{-4}$ per cycle, see Fig.~\ref{fig:error_budget}(c).

\begin{table}[h]
    \setlength{\tabcolsep}{12pt}
    \begin{tabular}{ccc}
    Transition   & Transition frequency & ac Zeeman shift \\
    \hline\rule{0pt}{2.5ex}$\ket{4,+4} \Leftrightarrow \ket{3,+3}$   &  2.547~GHz & $3.37(2)$~kHz \\
    $\ket{4, +2} \Leftrightarrow \ket{3,+3}$   & 2.791~GHz & $4.35(4)$~kHz \\
    $\ket{4, +2} \Leftrightarrow \ket{3, +2}$   & 2.908~GHz & $5.03(2)$~kHz \\
    \hline \\
    \end{tabular}
    \caption{
    \textbf{ac Zeeman shifts induced by MW $\pi$-pulses.}
    The transitions used to transfer population within the ground state are listed with their frequencies and the measured ac Zeeman shifts they induce on the data qubit.
    }
    \label{tab:transfer_pulses}
\end{table}

\subsection{Error induced by laser pulses}
Laser pulses at 729~nm, 854~nm, and 850~nm are required to re-initialise the coolant ion after the sideband pulse is applied (described in Sec.~\ref{SI:scheme_overview}), which lead to two types of errors.
Firstly, the 729~nm $\pi$-pulses induce an ac Stark shift of $12(1)$~Hz on the data qubit.
However, by applying the spin-echo scheme described above, the error induced by the laser pulses is reduced to an amount indistinguishable from the decoherence that would occur on the qubit for delays of the same duration as the laser pulses, see Fig.~\ref{fig:error_budget}(d).
Secondly, when the population in the $4\text{P}_{3/2}$ manifold decays back to the $4\text{S}_{1/2}$ manifold, a 393~nm photon is emitted which may be absorbed by the data qubit.
Through simulation, we estimate this to contribute an error of $1.6(9)\times 10^{-6}$ per cooling cycle, assuming an ion-to-ion distance of $5~\upmu$m and a uniform emission pattern.
Unlike the technical errors caused by the sideband and MW $\pi$-pulses, this is a fundamental error which cannot be corrected.
A possible way of improving this error is by choosing a different cooling transition with a greater frequency separation from the data qubit transition, see Sec.~\ref{SI:cooling_transition}, such that the emitted 393~nm photon has a larger frequency detuning from any transition between the data qubit states, $\ket{4\text{S}_{1/2}, 4, +1}$ and $\ket{4\text{S}_{1/2}, 3, +1}$, and states in the $\text{P}_{3/2}$ manifold.

\newpage

\section{Microwave drive chain}~\label{SI:mw_chain}

\begin{figure*}[h!]
    \centering
    \includegraphics[width=0.9\textwidth]{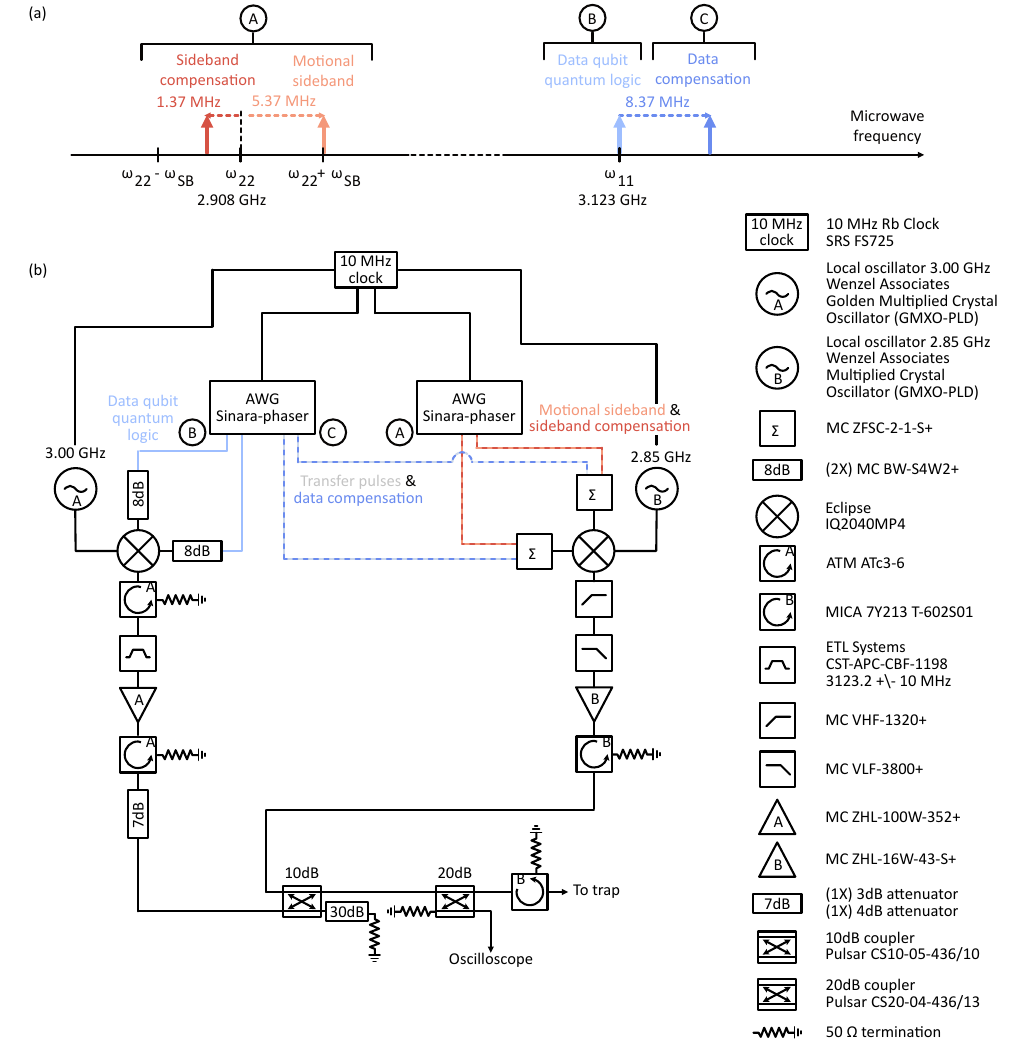}
    \caption{
    \textbf{Microwave pulse generation.}
    \textbf{(a)} An overview of the MW tones used for MW sideband cooling whilst protecting the data qubit.
    The motional sideband tone (orange) and the sideband compensation tone (red) are detuned from the coolant transition $\ket{3, 2}$ to $\ket{4, 2}$.
    The data compensation tone (blue) is detuned from the data transition $\ket{3, 1}$ to $\ket{4, 1}$ and the data qubit quantum logic tone (grey) resonantly drives the data transition.
    In addition, MW transfer pulses at frequencies of 2.547 to 3.123~GHz are used to transfer population around the ground state manifold.
    \textbf{(c)} Schematic of the MW drive chain.
    The circuitry on the left produces pulses at only the data qubit frequency (3.123~GHz) and the circuitry on the right produces pulses in a wider range of frequencies.
    The two chains are combined before being delivered to the on-chip MW resonator via a MW vacuum feedthrough.
    Three arbitrary waveform generator (AWG) channels are used to produce all of the required MW tones: (A) motional sideband and sideband compensation tones centred around the $\ket{4, 2}$ to $\ket{3, 2}$ transition, (B) data qubit quantum logic operations -- such as the spin-echo pulses and the Clifford gates used for benchmarking the error induced on the data qubit -- and (C) transfer pulses and the data compensation tone.
    }
    \label{fig:mw_chain}
\end{figure*}

In Fig.~\ref{fig:mw_chain}, we show the microwave setup used to drive operations on both coolant and data ions.

\end{document}